# Resonant Ultrasound Spectroscopy: Asymptotic Behavior of Resonant Frequencies

Farhad Farzbod,[1] and Casey Holycross[2]

[1] *Department of Mechanical Engineering, University of Mississippi, University, MS 38677, USA*

[2] *Aerospace Systems Directorate (AFRL/RQTI), Wright-Patterson AFB, OH 45433, USA*

Resonance ultrasound spectroscopy (RUS) is a non-destructive technique used to assess materials' elastic and anelastic properties. It involves measuring the frequencies of free vibrations in a carefully prepared sample to extract material properties. In this paper, we investigate the asymptotic behavior of eigenfrequencies. Our primary focus is on analyzing the asymptotic behavior of eigenfrequencies, aiming to understand their rate of growth and convergence. We also make observations regarding the impact of elastic constants on eigenfrequencies.

## I. INTRODUCTION

Resonance ultrasound spectroscopy (RUS) is a non-destructive method for evaluating materials' properties by measuring the resonant vibrations of a sample. RUS has been utilized since its inception to assess a wide range of material characteristics. These include pioneering works to find elastic constants [1-7], piezoelectric coefficients [6], crystallographic orientation [8, 9], and dislocation density [10]. With the widespread adoption of additive manufacturing techniques in the industry, RUS has emerged as a valuable tool for detecting the microstructural effects that arise from heat treatment in 3-D printed metals [11]. Additionally, RUS has proven effective in controlling the quality of complex additively manufactured components [12]. In conventional RUS, piezoelectric transducers have traditionally been employed for both the excitation and detection of ultrasound [1, 3, 4, 13]. However, to reduce the coupling between transducers and the sample, alternative non-contact transducer methods have been explored for RUS. These include the utilization of electromagnetic acoustic transducers [14] and lasers [15-23]. Such non-contact transducers have been employed to achieve effective RUS measurements while minimizing direct physical contact with the sample. The flexibility of this technique extends to the measurement of elastic properties in diverse and unique materials like cement building materials [24, 25], materials with residual stress [26], initial strains [27], human dentin [28], and human bone [29]. Additionally, RUS can be applied to various geometries, boundary conditions, and temperatures. For instance, it can accommodate cantilever boundary conditions [30], thin layer configurations [27], as well as elevated temperatures [28, 29].

In RUS, material properties are extracted by comparing the measured resonant peaks with a theoretically calculated spectrum using an error function. To minimize the error, the inputs to the spectral estimation are iteratively adjusted. Traditionally, the Levenberg-Marquardt algorithm has been employed for back-calculating the material properties. Other techniques, such as Bayesian inference or multinomial classification, have also been recently utilized to perform error analysis of the results in estimating the material properties.

The organization of the paper is as follows: Firstly, we provide a review of the mathematical background for calculating resonant modes of free vibrations. Secondly, we delve into the investigation of the asymptotic behavior of the resonant frequencies. Next, we analyze and compare the numerical resonant frequencies with the results of the theoretical analysis. Finally, we discuss the valuable insights obtained from studying the asymptotic behavior and the influence of elastic constants.

## II. ASYMPTOTIC BEHAVIOR OF RESONANT FREQUENCIES

### A. RUS Calculation: Background

The comprehensive methodology for computing resonant frequencies in an elastic solid is extensively detailed in the works of Demarest [6] and Visscher et. Al. [31]. These papers provide in-depth explanations of the generalized approach, which involves formulating the Lagrangian and determining the frequencies and mode shapes that leads to the extremum value of the Lagrangian. This approach allows us to obtain a solution to the elastic wave equation, facilitating the computation of resonant frequencies in the elastic solid.

The Lagrangian for the system is expressed as Eq. (1), which encompasses the kinetic and potential energy terms as presented in Eq. (2). In this formulation, the displacement is assumed to be harmonic, with the frequency denoted as $\omega$, and described by the variables of mass density $\varrho$ and displacement u. The Einstein convention is utilized, where the indices i, j, k, l range from 1 to 3, representing the three spatial dimensions.

$$L = \frac{1}{2}\int_V \left(\rho\omega^2 u_i^2 - C_{ijkl}\frac{\partial u_i}{\partial x_j}.\frac{\partial u_k}{\partial x_l}\right) dV \tag{1}$$

$$KE = \frac{1}{2}\ \rho\omega^2 u_i^2 \qquad PE = \frac{1}{2}C_{ijkl}\frac{\partial u_i}{\partial x_j}.\frac{\partial u_k}{\partial x_l} \tag{2}$$

To obtain an approximation of the displacements, the Rayleigh-Ritz method is utilized, employing a finite functional basis. In this method, the displacements in the *i* direction, denoted as $u_i$ are expressed using chosen basis functions $\varphi_q$:

$$u_i = a_{iq}\varphi_q \tag{3}$$

By expanding and rearranging the Lagrangian, we arrive at Eq. (4). In this form, the volume integrals can be evaluated independently of the coefficients $a_{iq}$. Therefore, we factor out these terms to simplify the expression.

$$L = \frac{1}{2}a_{iq}a_{i'q'}\rho\omega^2 \int_V \delta_{ii'}\varphi_q(x)\varphi_{q'}(x)dV - \frac{1}{2}a_{iq}a_{kq'}\int_V C_{ijkl}\frac{\partial\varphi_q}{\partial x_j}.\frac{\partial\varphi_{q'}}{\partial x_l}dV \tag{4}$$

The calculation of volume integrals involving basis functions, their derivatives, and $C_{ijkl}$ can be performed independently of $a_{ij}$. We can represent these matrices as **E** and **Γ**, respectively. Eq. (4) can be expressed as follows:

$$L = \frac{1}{2}(\rho\omega^2 \boldsymbol{a}^T\mathbf{E}\boldsymbol{a} - \boldsymbol{a}^T\boldsymbol{\Gamma}\boldsymbol{a}) \tag{5}$$

where the coefficients $a_{iq}$ are expressed as vectors. Our objective is to determine the values of $\boldsymbol{a}$ that extremize the value of *L*. By employing the Rayleigh-Ritz method, we set the derivative of *L* with respect to $\boldsymbol{a}$ equal to zero:

$$\rho\omega^2\mathbf{E}\boldsymbol{a} - \boldsymbol{\Gamma}\boldsymbol{a} = 0 \tag{6}$$

This equation represents a general eigenvalue problem. There are various options for the basis functions. In the next subsection, we use Legendre's polynomials.

### B. Asymptotic Analysis

In this study, we focus on cuboid samples positioned with their volume center coinciding with the origin of the coordinate system. For these cases, the Legendre polynomials are selected as the basis functions. These polynomials are employed to represent $\varphi_q$ in Eq. (3) as a function of the spatial variables $x_1$, $x_2$ and $x_3$ (along the *x*, *y*, and *z* axes). The index *q* in $\varphi_q$ signifies a set of three non-negative integers (*k*,*l*,*m*) that correspond to the degrees of the Legendre polynomials. As such $\varphi_q$ is represented by:

$$\varphi_q(\mathrm{x}) = \left(\frac{P_k\left(\frac{2x_1}{D_1}\right)}{\sqrt{\frac{D_1}{2k+1}}}\right)\left(\frac{P_l\left(\frac{2x_2}{D_2}\right)}{\sqrt{\frac{D_2}{2l+1}}}\right)\left(\frac{P_m\left(\frac{2x_3}{D_3}\right)}{\sqrt{\frac{D_3}{2m+1}}}\right),\ q = 1,..,n.\ ,\begin{cases} -\frac{D_1}{2} < x_1 < \frac{D_1}{2} \\ -\frac{D_2}{2} < x_2 < \frac{D_2}{2} \\ -\frac{D_3}{2} < x_3 < \frac{D_3}{2} \end{cases} \tag{7}$$

where $D_1$, $D_2$ and $D_3$ represent the dimensions of the sample along the *x*, *y* and *z* axes, respectively. The numerators within each parenthesis in Eq. (7) correspond to the standard Legendre polynomial expressed by Eq. (8), with the orthogonality property [32] stated in Eq. (9) where δ represents the Kronecker delta. Since the origin coincides with the center of the cuboid sample, we normalize the Legendre polynomials in Eq. (7) by dividing each Legendre polynomial by its norm over the corresponding interval along $x_1$, $x_2$ and $x_3$, as described in Eq. (10).

In Eq. (7), the dimension *n* of the basis is restricted by the value of N, as defined in Eq. (10), where N is determined by Eq. (11) and *n* is determined by Eq. (12). Typically, N = 10 is selected, as it is considered suitable for approximation of the first 50 eigenmodes [1]. However, in this study, we are investigating asymptotic behavior and *n* approaches infinity.

$$P_k(x) = \frac{1}{2^k k!}\frac{d^k}{dx^k}(x^2-1)^k \tag{8}$$

$$\int_{-1}^{1} P_k(x)P_l(x)dx = \frac{2\delta_{kl}}{2k+1} \tag{9}$$

$$\left\| P_k\left(\frac{2x_1}{D_1}\right) \right\| = \sqrt{\int_{-\frac{D_1}{2}}^{\frac{D_1}{2}} \left(P_k\left(\frac{2x_1}{D_1}\right)\right)^2 dx_1} = \sqrt{\frac{D_1}{2k+1}} \tag{10}$$

$$k + l + m \leq N \tag{11}$$

$$n = \frac{3(N+1)(N+2)(N+3)}{6} \tag{12}$$

Because of our choice of basis function and the orthonormality relation, the first integral in Eq. (4) vanishes for q≠q' and equals to 1 otherwise. Consequently, the matrix E becomes a unity matrix and Eq. (6) becomes a standard eigenvalue problem:

$$\mathbf{\Gamma}\boldsymbol{a} = \lambda\boldsymbol{a} \tag{13}$$

In order to simplify the equation, we replace $\varrho\omega^2$ with $\lambda$. In certain sections of this paper, we work with the eigenvalue $\lambda$ rather than the eigenfrequency ω. Our main objective is to determine the asymptotic behavior of $\lambda$ in Eq. (13). To achieve this, we proceed with a series of steps.

### *1. Adding more elements leads to higher eigenfrequencies*

The number of eigenvalues in Eq. (13) is the same as the number of elements *n*. It is a widely accepted principle in engineering and physics that increasing the number of elements results in higher eigenfrequencies in a monotonic manner. In other words, if we have *n*+1 elements, we would have *n*+1 eigenvalues, and for the new eigenvalues, we have $\lambda_{n+1} > \lambda_n$. The literature offers substantial evidence supporting this concept, and its proof is well-established. However, for a thorough understanding of asymptotic behavior and to facilitate a comprehensive discussion, it is beneficial to delve into this topic. Detailed explanations and proofs of this step, along with a comprehensive exploration of the Rayleigh-Ritz method, can be found in references [33-35].

By multiplying both sides of the equation by $a^T$, we obtain the Rayleigh quotient:

$$\lambda = \frac{\boldsymbol{a}^T\mathbf{\Gamma}\boldsymbol{a}}{\boldsymbol{a}^T\boldsymbol{a}} \tag{14}$$

When considering the use of *n* basis functions in Eq. (3), where *q* ranges from 1 to *n* for $\varphi_q$ the matrix $\mathbf{\Gamma}$ would have dimensions of $3n\times3n$. In this case, we can define a function:

$$R: \mathbb{R}^n \to \mathbb{R},\ R(\boldsymbol{a}) = \frac{\boldsymbol{a}^T\mathbf{\Gamma}\boldsymbol{a}}{\boldsymbol{a}^T\boldsymbol{a}} \tag{15}$$

This function, R(a), reaches its minimum value when the input vector $\boldsymbol{a}$ is the first eigenvector of $\mathbf{\Gamma}$, denoted as $\boldsymbol{a}_{\lambda_1}$ (or its multiples), and $R(\boldsymbol{a})$ equals its first eigenvalue, $\lambda_1$. It is worth noting that the matrix $\mathbf{\Gamma}$, as implied by Eq. (4), is symmetric (as it represents a stiffness matrix), ensuring that its eigenvectors are orthogonal. Furthermore, if we consider the subspace of $\mathbb{R}^n$ that is orthogonal to $\boldsymbol{a}_{\lambda_1}$, which is known as the orthogonal complement subspace of $\boldsymbol{a}_{\lambda_1}$, we find that in this subspace, $\boldsymbol{a}_{\lambda_2}$ minimizes the Rayleigh quotient and assumes the value of $\lambda_2$, the second eigenvalue. This argument holds true for the remaining eigenvalues and eigenvectors of $\boldsymbol{\Gamma}$, of which there are $3n$ in this particular case. Consequently, the eigenvalues follow an increasing order, meaning that $\lambda_1 < \lambda_2 < \lambda_3$.... This implies that the eigenvalues become larger as we move along the sequence.

Furthermore, when we solve Eq. (13) using a basis comprising a finite number of functions (denoted by $n$), increasing the value of $n$ expands the subspace within which the minimization process takes place. Consequently, the minimum value of the Rayleigh quotient decreases as the dimension of the subspace increases. For example, if we use the basis functions $P_1$, $P_2$, …$P_{50}$, the tenth eigenvalue would be smaller compared to the case when we use $P_1$, $P_2$, …$P_{20}$ as the basis functions in Eq. (13). In practical terms, if we are interested in the first hundred resonant frequencies, we observe that beyond a particular dimension, the reduction in eigenvalues becomes less significant and falls below a desired error threshold.

Additionally, it can be inferred that the approximate eigenfrequencies obtained by using finite-dimensional basis functions to solve Eq. (13) are always greater than the resonant frequencies of Eq. (1).

### *2. Some key properties of Legendre's polynomials*

The following equation, Eq. (16), establishes a recurrence relation between the derivatives of Legendre's polynomials. This relation can be found in textbooks such as [32, 36]:

$$\frac{d}{dx}P_{k+1}(x) - \frac{d}{dx}P_{k-1}(x) = (2k+1)P_k(x) \tag{16}$$

We can use Eq. (16) for all $P_i$'s from $k$ down by increments of two, resulting in the following set of equations:

$$\begin{cases} \frac{d}{dx}P_k(x) - \frac{d}{dx}P_{k-2}(x) = (2(k-1)+1)P_{k-1}(x) \\ \frac{d}{dx}P_{k-2}(x) - \frac{d}{dx}P_{k-4}(x) = (2(k-3)+1)P_{k-3}(x) \\ \frac{d}{dx}P_{k-4}(x) - \frac{d}{dx}P_{k-6}(x) = (2(k-5)+1)P_{k-5}(x) \\ \qquad\qquad\qquad \vdots \end{cases} \tag{17}$$

For the right-hand side of these equations, we can utilize Eq. (9) to express the norms of Legendre's polynomials. This allows us to rewrite them as follows:

$$(2(k-1)+1)P_{k-1}(x) = \frac{2P_{k-1}(x)}{\|P_{k-1}(\mathrm{x})\|^2},\ (2(k-3)+1)P_{k-3}(x) = \frac{2P_{k-3}(x)}{\|P_{k-3}(\mathrm{x})\|^2}, \ldots \tag{18}$$

Considering the set of equations in Eqs. (17), we can add them all together. While Eqs. (17) includes only the first three equations, the remaining equations are represented by three dots. On the left-hand side, all the terms cancel out except for the first term of the first equation, which is $\frac{d}{dx}P_k(x)$, and the second term of the last equation. This second term of the last equation in Eqs. (17) can be either $-\frac{d}{dx}P_0(x)$ or $-\frac{d}{dx}P_1(x)$, depending on whether $k$ is even or odd, respectively. For the case of even $k$, going down by increments of two results in the second term of the last equation in Eq. (17) being $-\frac{d}{dx}P_0(x)$. However, since $P_0(x)$ is equal to 1, the term $-\frac{d}{dx}P_0(x)$ is zero and can be neglected. The sum on the right-hand side goes from $k$-1 down to 1 with increments of two. Therefore, we can write $\frac{d}{dx}P_k(x)$ as the sum of Legendre's polynomials for the case of even $k$.

For the case of odd $k$, the sum on the right-hand side goes from $k$-1 down to 2 with increments of two. The last term on the left-hand side, $-\frac{d}{dx}P_1(x)$, equals -2, which is equivalent to $-\frac{2P_0(x)}{\|P_0(x)\|^2}$. To simplify, we can add 2 to both sides of the equation, allowing the sum on the left-hand side to go from $k$-1 down to zero. Hence, for both odd and even cases of $k$, we can express $\frac{d}{dx}P_k(x)$ as the sum of Legendre's polynomials:

$$\frac{d}{dx}P_k(x) = \frac{2P_{k-1}(x)}{\|P_{k-1}(x)\|^2} + \frac{2P_{k-3}(x)}{\|P_{k-3}(x)\|^2} + \frac{2P_{k-5}(x)}{\|P_{k-5}(x)\|^2} + \cdots \quad (19)$$

### 3. *Trace of the matrix Γ*

In this subsection, we aim to derive the formulation for the trace of the matrix $\mathbf{\Gamma}$. By expressing the displacement in the $i$ direction according to Eq. (30), the potential energy term $\boldsymbol{a}^T\mathbf{\Gamma}\boldsymbol{a}$ in Eqs. (4-5) can be written as:

$$\boldsymbol{a}^T\mathbf{\Gamma}\boldsymbol{a} = \begin{bmatrix}\boldsymbol{a_1} & \boldsymbol{a_2} & \boldsymbol{a_3}\end{bmatrix}\begin{bmatrix}\mathbf{\Gamma_{11}} & \mathbf{\Gamma_{12}} & \mathbf{\Gamma_{13}} \\ \mathbf{\Gamma_{12}} & \mathbf{\Gamma_{22}} & \mathbf{\Gamma_{23}} \\ \mathbf{\Gamma_{13}} & \mathbf{\Gamma_{23}} & \mathbf{\Gamma_{33}}\end{bmatrix}\begin{bmatrix}\boldsymbol{a_1} \\ \boldsymbol{a_2} \\ \boldsymbol{a_3}\end{bmatrix}, \quad (20)$$

where:

$$\boldsymbol{a_i} = \begin{bmatrix}a_{i1} & a_{i2} & a_{i3} & a_{i4} \dots & a_{in}\end{bmatrix}, \quad (21)$$

and

$$\begin{cases}\mathbf{\Gamma_{11}} = \mathrm{C}_{11}\mathbf{\Phi_{11}} + \mathrm{C}_{61}\mathbf{\Phi_{21}} + \mathrm{C}_{51}\mathbf{\Phi_{31}} + \mathrm{C}_{16}\mathbf{\Phi_{12}} + \mathrm{C}_{66}\mathbf{\Phi_{22}} + \mathrm{C}_{56}\mathbf{\Phi_{32}} + \mathrm{C}_{15}\mathbf{\Phi_{13}} + \mathrm{C}_{65}\mathbf{\Phi_{23}} + \mathrm{C}_{55}\mathbf{\Phi_{33}} \\ \mathbf{\Gamma_{22}} = \mathrm{C}_{66}\mathbf{\Phi_{11}} + \mathrm{C}_{26}\mathbf{\Phi_{21}} + \mathrm{C}_{46}\mathbf{\Phi_{31}} + \mathrm{C}_{62}\mathbf{\Phi_{12}} + \mathrm{C}_{22}\mathbf{\Phi_{22}} + \mathrm{C}_{42}\mathbf{\Phi_{32}} + \mathrm{C}_{64}\mathbf{\Phi_{13}} + \mathrm{C}_{24}\mathbf{\Phi_{23}} + \mathrm{C}_{44}\mathbf{\Phi_{33}} \\ \mathbf{\Gamma_{33}} = \mathrm{C}_{55}\mathbf{\Phi_{11}} + \mathrm{C}_{45}\mathbf{\Phi_{21}} + \mathrm{C}_{35}\mathbf{\Phi_{31}} + \mathrm{C}_{54}\mathbf{\Phi_{12}} + \mathrm{C}_{44}\mathbf{\Phi_{22}} + \mathrm{C}_{34}\mathbf{\Phi_{32}} + \mathrm{C}_{53}\mathbf{\Phi_{13}} + \mathrm{C}_{43}\mathbf{\Phi_{23}} + \mathrm{C}_{33}\mathbf{\Phi_{33}}\end{cases} \quad (22)$$

In the above Eq. (22) $C_{ij}$ represents the elastic constants in Voigt notation and the matrices $\Phi_{ij}$ are defined as:

$$\mathbf{\Phi_{ij}} = \begin{bmatrix}\int_V \varphi_{1,i}\ \varphi_{1,j}\ dV & \int_V \varphi_{1,i}\ \varphi_{2,j}\ dV & \int_V \varphi_{1,i}\ \varphi_{3,j}\ dV & \cdots & \int_V \varphi_{1,i}\ \varphi_{s,j}\ dV & \cdots \\ \int_V \varphi_{2,i}\ \varphi_{1,j}\ dV & \int_V \varphi_{2,i}\ \varphi_{2,j}\ dV & \int_V \varphi_{2,i}\ \varphi_{3,j}\ dV & \cdots & \int_V \varphi_{2,i}\ \varphi_{s,j}\ dV & \cdots \\ \vdots & \vdots & \vdots & \vdots & \vdots & \vdots \\ \int_V \varphi_{r,i}\ \varphi_{1,j}\ dV & \int_V \varphi_{r,i}\ \varphi_{2,j}\ dV & \int_V \varphi_{r,i}\ \varphi_{3,j}\ dV & \cdots & \int_V \varphi_{r,i}\ \varphi_{s,j}\ dV & \cdots \\ \vdots & \vdots & \vdots & \vdots & \vdots & \vdots\end{bmatrix} \quad (23)$$

The $\varphi_r$'s represent the basis functions as defined in Eq. (7), and the comma represents the derivative, i.e., $\varphi_{r,i} = \frac{\partial\varphi_r}{\partial x_i}$. Additionally, we have $\mathbf{\Phi_{ij}} = \mathbf{\Phi_{ji}}^{\boldsymbol{T}}$ which can be utilized to simplify the computation of Eq. (22). However, this property has been discussed in numerous previous works and is not a focus of our current analysis..

Our intermediate objective is to analyze the sum of the eigenvalues of $\mathbf{\Gamma}$ in Eq. (13), which is equivalent to the trace of $\mathbf{\Gamma}$. It can be observed from Eq. (22) that to calculate the trace of $\mathbf{\Gamma}$, we need to compute the trace of the $\mathbf{\Phi}_{ij}$ matrices in Eq. (23).

### 4. *Analyzing the trace of the matrix Γ*

In our analysis, we aim to investigate the asymptotic behavior of eigenvalues by examining the sum of eigenvalues, which is equivalent to the trace of $\mathbf{\Gamma}$. This sum is determined by the traces of $\mathbf{\Gamma}_{11}$, $\mathbf{\Gamma}_{22}$ and $\mathbf{\Gamma}_{33}$. Referring to Eq. (22), these traces depend on the elastic constants and the traces of the $\mathbf{\Phi}_{ij}$ matrices. The traces of the $\mathbf{\Phi}_{ij}$ matrices consists of elements in the form:

$$\int_V \varphi_{q,i}\ \varphi_{q,j}\ dV \tag{24}$$

Firstly, we will analyze the situations where *i≠j*. To simplify the discussion, we can assume, without loss of generality, that *i*=1 and *j*=2. Under this assumption, Eq (24) can be rewritten as:

$$\int_V \left(\frac{\frac{d}{dx_1}P_k\left(\frac{2x_1}{D_1}\right)}{\sqrt{\frac{D_1}{2k+1}}}\right)\left(\frac{P_l\left(\frac{2x_2}{D_2}\right)}{\sqrt{\frac{D_2}{2l+1}}}\right)\left(\frac{P_m\left(\frac{2x_3}{D_3}\right)}{\sqrt{\frac{D_3}{2m+1}}}\right)\left(\frac{P_k\left(\frac{2x_1}{D_1}\right)}{\sqrt{\frac{D_1}{2k+1}}}\right)\left(\frac{\frac{d}{dx_2}P_l\left(\frac{2x_2}{D_2}\right)}{\sqrt{\frac{D_2}{2l+1}}}\right)\left(\frac{P_m\left(\frac{2x_3}{D_3}\right)}{\sqrt{\frac{D_3}{2m+1}}}\right)dV \tag{25}$$

Now, considering the first term in the integral, the k[th] Legendre's polynomial can be expressed in terms of lower-order Legendre polynomials, as indicated by Eq. (19). As a result, the derivative of the k[th] term is orthogonal to the k[th] term itself leading to the integral vanishing. This holds true for all integral terms on the diagonal of $\mathbf{\Phi}_{ij}$ for *i≠j*, effectively making their trace equal to zero. However, when *i=j,* the situation is different as a derivative is multiplied by itself in the integral. The trace of $\mathbf{\Phi}_{ii}$ can be calculated as:

$$\sum_{q=1}^{\frac{(N+1)(N+2)(N+3)}{6}} \int_V \varphi_{q,i}\ \varphi_{q,i}\ dV. \tag{26}$$

To compute the integral within the summation for each basis function, let's begin by determining the integral for the normalized Legendre polynomial over the interval [-1, 1]. The normalized Legendre polynomial can be expressed as follows:

$$L_k(x) = \frac{P_k(x)}{\|P_k(\mathrm{x})\|} = \sqrt{\frac{2k+1}{2}}P_k(x) \tag{27}$$

Next, we can utilize Eq. (19) for the derivative, which leads to the following expression:

$$\int_{-1}^{1}\left(\frac{d}{dx}P_k(x)\right)^2 dx = \int_{-1}^{1}\left(\frac{2P_{k-1}(x)}{\|P_{k-1}(\mathrm{x})\|^2} + \frac{2P_{k-3}(x)}{\|P_{k-3}(\mathrm{x})\|^2} + \frac{2P_{k-5}(x)}{\|P_{k-5}(\mathrm{x})\|^2} + \cdots\right)^2 dx \tag{28}$$

Due to the orthogonality of Legendre polynomials, the terms with cross multiplication in the right-hand side integral of Eq. (28) vanish, resulting in the simplification of the right-hand side as follows:

$$\int_{-1}^{1}\left[\left(\frac{2P_{k-1}(x)}{\|P_{k-1}(\mathrm{x})\|^2}\right)^2 + \left(\frac{2P_{k-3}(x)}{\|P_{k-3}(\mathrm{x})\|^2}\right)^2 + \left(\frac{2P_{k-5}(x)}{\|P_{k-5}(\mathrm{x})\|^2}\right)^2 + \cdots\right]dx \tag{29}$$

Since the integral of each squared Legendre polynomial equals the square of its norm, Eq. (29) can be further simplified as:

$$\frac{4}{\|P_{k-1}(\mathrm{x})\|^2} + \frac{4}{\|P_{k-3}(\mathrm{x})\|^2} + \frac{4}{\|P_{k-5}(\mathrm{x})\|^2} + \cdots = 4[(k-1) + (k-3) + (k-5) + \cdots] + 2(1+1+1+..) \tag{30}$$

Now, we can consider two different cases for the above sums: when *k* is even and when *k* is odd.:

$$k\ even\!: \begin{cases} (k-1) + (k-3) + (k-5) + \cdots = \frac{k^2}{4} \\ (1+1+\cdots) = \frac{k}{2} \end{cases},\ k\ odd\!: \begin{cases} (k-1) + (k-3) + (k-5) + \cdots = \frac{(k-1)(k+1)}{4} \\ (1+1+\cdots) = \frac{k+1}{2} \end{cases} \tag{31}$$

In both cases, the summation in Eq. (30) simplifies to the same expression, namely *k(k+1).* We can utilize Eq. (27) to rewrite Eq. (28) as follows:

$$\int_{-1}^{1}\left(\frac{d}{dx}L_k(x)\right)^2 dx = \frac{k(k+1)(2k+1)}{2} \tag{32}$$

Consequently, by introducing a change of variables, we obtain the following:

$$\int_{-\frac{D_i}{2}}^{\frac{D_i}{2}}\left(\frac{\frac{d}{dx_1}P_k\left(\frac{2x_1}{D_1}\right)}{\sqrt{\frac{D_1}{2k+1}}}\right)^2 dx_i = \int_{-\frac{D_i}{2}}^{\frac{D_i}{2}}\left(\frac{d}{dx_i}L_k\left(\frac{2x_\mathrm{i}}{D_\mathrm{i}}\right)\right)^2 dx_i = \frac{2k(k+1)(2k+1)}{{D_i}^2}, \qquad i = 1, 2, 3 \tag{33}$$

It is worth mentioning that although this equation, along with similar ones for other derivatives, is not the focus of our current work, it can be utilized to expedite calculations related to RUS when employing Legendre polynomials.

Let's consider that we have initially selected N, as defined in Eq. (11), to be a specific number. This results in n eigenvalues, where *n* is defined in Eq. (12). Now if we increment N by 1, the number of eigenvalues increases by:

$$\frac{3(N+2)(N+3)(N+4)}{6} - \frac{3(N+1)(N+2)(N+3)}{6} = \frac{3N^2+15N+18}{2} \tag{34}$$

To calculate the sum of the added eigenvalues, we can examine the change in the trace of **Γ**. For this calculation, we employ Eq. (33). The added elements to the trace correspond to those elements in which the powers of $x_1$, $x_2$, $x_3$, in the Legendre polynomial of the basis functions are increased by N+1. In other words, $k + l + m =$ N+1 in Eq. (7). The additional value contributed to the trace can be expressed as:

$$\sum_{i=1,2,3}\sum_{q=\frac{(N+1)(N+2)(N+3)}{6}}^{\frac{(N+2)(N+3)(N+4)}{6}}\int_V \varphi_{q,i}\ \varphi_{q,i}\ dV. \tag{35}$$

Let's first consider the case for $i$=1 within the inner sigma. In this scenario, we have:

$$\int_V \varphi_{q,1}\ \varphi_{q,1}\ dV = \int_V \left(\frac{\frac{d}{dx_1}P_k\left(\frac{2x_1}{D_1}\right)}{\sqrt{\frac{D_1}{2k+1}}}\right)^2 \left(\frac{P_l\left(\frac{2x_2}{D_2}\right)}{\sqrt{\frac{D_2}{2l+1}}}\right)^2 \left(\frac{P_m\left(\frac{2x_3}{D_3}\right)}{\sqrt{\frac{D_3}{2m+1}}}\right)^2 dV \tag{36}$$

In Eq. (36), the integrals of the second and third terms on the right-hand side, over the $x_2$ and $x_3$ axes of the volume integral, respectively, both evaluate to one. The remaining integral of the first term can be calculated using Eq. (33), yielding:

$$\int_V \varphi_{q,1}\ \varphi_{q,1}\ dV = \frac{2k(k+1)(2k+1)}{{D_1}^2}, \tag{37}$$

Here, $q$ enumerates the triplets ($k,l,m$). To calculate the sum in Eq. (35) over $q$, we need to determine how many $q$ correspond to the same $k$, or in other words, all the possible combinations for which $k+l+m$ = N+1.

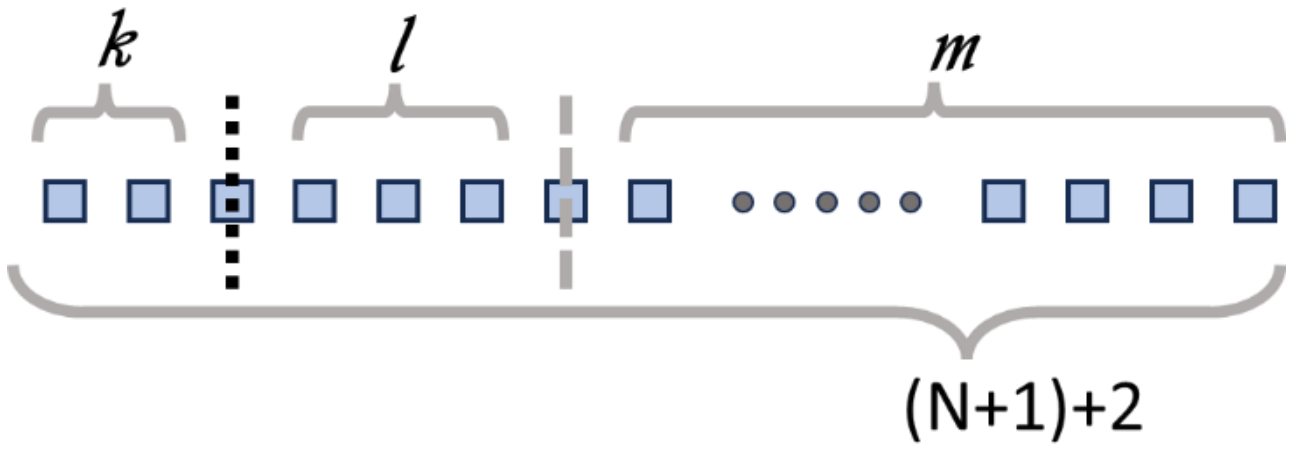

Figure 1: To determine all the possible cases for $q$ with the same $k$ and $k+l+m$=N+1, we consider (N+1)+2 spots. Among these spots, two of them will be crossed out to divide the spots into three regions representing ($k$, $l$, $m$). When we cross out one spot using a black dotted line to select the number k to the left of it, there will remain (N+3-1-k) choices for the spot of the gray dashed line.

To determine the combinations, as depicted in Fig. (1), we envision N+3 spots. Among these spots, two will be crossed out, and the remaining ones represent ($k,l,m$). If we cross out one spot using a black dotted line to select the number k to the left of it, then for the gray dashed line, we have (N+3-1-$k$) choices. This represents the number of different combinations of $q$ with the condition of $k+l+m$ = N+1. Therefore, the sum of Eq. (35) for $i$=1 as $k$ runs from 0 to N+1, can be expressed as follows:

$$\sum_{k=0}^{N+1}\frac{1}{{D_1}^2}2k(k+1)(2k+1)(N+3-1-k)=\frac{1}{{D_1}^2}\left[(4+2N)\sum_{k=0}^{N+1}k+(10+6N)\sum_{k=0}^{N+1}k^2+(2+4N)\sum_{k=0}^{N+1}k^3-4\sum_{k=0}^{N+1}k^4\right]. \qquad (38)$$

By utilizing formulas for the sum of natural numbers, their squares, the third power, and the fourth power, we can rewrite Eq. (38) as:

$$\frac{1}{{D_1}^2}\left[\frac{(4+2N)(N+1)(N+2)}{2}+\frac{(10+6N)(N+1)(N+2)(2N+3)}{6}+\frac{(2+4N)(N+1)^2(N+2)^2}{4}-4\left(\frac{6(N+1)^5+15(N+1)^4+10(N+1)^3-(N+1)}{30}\right)\right]=\frac{1}{{D_1}^2}\left[\frac{N^5}{5}+\frac{5N^4}{2}+\frac{12N^3}{1}+\frac{55N^2}{2}+\frac{149N}{5}+12\right]. \qquad (39)$$

A similar procedure can be applied for the cases of $i$ = 2 and $i$ = 3, resulting in equations analogous to Eq. (39) with $D_2$ and $D_3$ in the denominators. Considering Eqs. (22) for $\mathbf{\Gamma}_{11}$, $\mathbf{\Gamma}_{22}$ and $\mathbf{\Gamma}_{33}$, all three have terms with elastic constants and $\mathbf{\Phi}_{ii}$ similar to Eq. (39). Therefore, we can conclude that the added value to the trace of $\mathbf{\Gamma}$ when N is incremented to N+1 can be stated as:

$$b_5N^5+b_4N^4+b_3N^3+b_2N^2+b_1N+b_0 \qquad (40)$$

Here, the coefficients $b_i$'s are comprised of diagonal elements of the elastic constants from Eqs. (22) and numbers and dimensions from Eq. (39).

### *5. Asymptotes of the eigenvalues*

Regarding the eigenvalues, as discussed before, adding more elements results in higher eigenfrequencies. When we select N, we end up with $n=\frac{3(N+1)(N+2)(N+3)}{6}=\frac{N^3+3N^2+11N+6}{2}$ eigenvalues. We refer to the highest eigenvalue as $\lambda_n$. Increasing N by one, according to Eq. (34), would add $\frac{3N^2+15N+18}{2}$ to the number of eigenvalues. Since the sum of all these added eigenvalues is equal to the added value to the trace of $\mathbf{\Gamma}$, and each of these added eigenvalues is greater than $\lambda_n$, we obtain:

$$\left(\frac{3N^2+15N+18}{2}\right)\lambda_n \leq b_5N^5 + b_4N^4 + b_3N^3 + b_2N^2 + b_1N + b_0 \tag{41}$$

On the other hand, when the chosen number changes from N-1 to N, $\lambda_n$ is greater than all the added eigenvalues. Hence, we have:

$$b_5(N-1)^5 + b_4(N-1)^4 + b_3(N-1)^3 + b_2(N-1)^2 + b_1(N-1) + b_0 \leq \left(\frac{3(N-1)^2+15(N-1)+18}{2}\right)\lambda_n \tag{42}$$

Expanding the polynomial in Eq. (42), all the coefficients of the fifth-degree polynomial of N would change except the one for the highest power of $N^5$. We denote these new coefficients as $e_i$. Combining Eq. (42) with Eq. (41) and dividing both sides by $n$, we get:

$$\frac{b_5N^5+e_4N^4+e_3N^3+e_2N^2+e_1N+e_0}{0.75N^5+3.75N^4+10.5N^3+59.25N^2+27.75N-13.5} \leq \frac{\lambda_n}{n} \leq \frac{b_5N^5+b_4N^4+b_3N^3+b_2N^2+b_1N+b_0}{0.75N^5+6N^4+24N^3+59.25N^2+72N+27} \tag{43}$$

Upon taking the limit of the left side and the right side of the inequality as N or $n$ approaches infinity, they both converge to the same value, equal to $4b_5/3$. This implies:

$$\lim_{n\to\infty}\frac{\lambda_n}{n} = \lim_{n\to\infty}\frac{\rho{\omega_n}^2}{n} = constant \tag{44}$$

A notable observation is that this constant depends on the diagonal elements of $\mathbf{\Gamma}$. As demonstrated in Eqs. (22) and subsection II-B-4, the diagonal elements of $\mathbf{\Gamma}$ relies on the diagonal elements of the elastic constant matrix. This sensitivity issue, is however, not the focus of this paper and has been previously investigated [37, 38].

## III. NUMERICAL RESULTS AND DISCUSSIONS

For the numerical calculations, we utilize elastic constants and densities of the materials, which were measured in previous studies. These values are tabulated in Table 1.

TABLE I. Material property used in numerical calculations.

| Material[a] | Crystal family (Number of elastic constants) | Elastic Constants (GPa) | Density (Kg/m$^3$) |
|---|---|---|---|
| Copper | Cubic (3) | $\begin{bmatrix} 168 & 121 & 121 & 0 & 0 & 0 \\ & 168 & 121 & 0 & 0 & 0 \\ & & 168 & 0 & 0 & 0 \\ & & & 75 & 0 & 0 \\ & & & & 75 & 0 \\ & & & & & 75 \end{bmatrix}$ | 8960 |
| Titanium Diboride | Hexagonal (5) | $\begin{bmatrix} 654 & 49 & 95 & 0 & 0 & 0 \\ & 654 & 49 & 0 & 0 & 0 \\ & & 458 & 0 & 0 & 0 \\ & & & 262 & 0 & 0 \\ & & & & 262 & 0 \\ & & & & & \frac{654-49}{2} \end{bmatrix}$ | 4520 |
| Chrome-Diopside | Monoclinic (13) | $\begin{bmatrix} 228.1 & 78.8 & 70.2 & 0 & 7.9 & 0 \\ & 181.1 & 61.1 & 0 & 5.9 & 0 \\ & & 245.4 & 0 & 39.7 & 0 \\ & & & 78.9 & 0 & 6.4 \\ & & & & 68.2 & 0 \\ & & & & & 78.1 \end{bmatrix}$ | 3286 |
| Silicon[b] | Resembling Triclinic (21) | $\begin{bmatrix} 197 & 56.86 & 39.81 & -6.52 & -6.75 & 4.11 \\ & 183.65 & 54.21 & 9.65 & 8.04 & -8.46 \\ & & 199.63 & -7.75 & -1.45 & 5.48 \\ & & & 70.23 & 4.46 & 8.18 \\ & & & & 55.58 & -5.46 \\ & & & & & 73.07 \end{bmatrix}$ | 2330 |

[a] These properties of copper, titanium diboride, chrome-diopside and silicon are from the references of [39], [40], [41] and [22], respectively.

[b] The elastic constants are for rotated single-crystal silicon.

## A. Numerical limits and convergence

In Eq. (44), $\lambda_n = \varrho\omega^2$, where ω represents the angular frequency. We plot $\frac{\lambda_n}{n}$ for the first sample of copper, and the dimensions $D_1$, $D_2$, and $D_3$ are equal to 1.5 mm, 2.5 mm, and 3.5 mm, respectively. As discussed in subsection II-B-1, the eigenvalues obtained using a finite basis are higher than the actual resonant frequencies. This observation is clearly evident in Fig. 2; with increasing N, the slope of the line approaches zero more rapidly and is sustained over a larger range. While assessing the accuracy of the RUS results using a finite basis is beyond the scope of this work, one might be able to employ this convergence criterion to evaluate the accuracy.

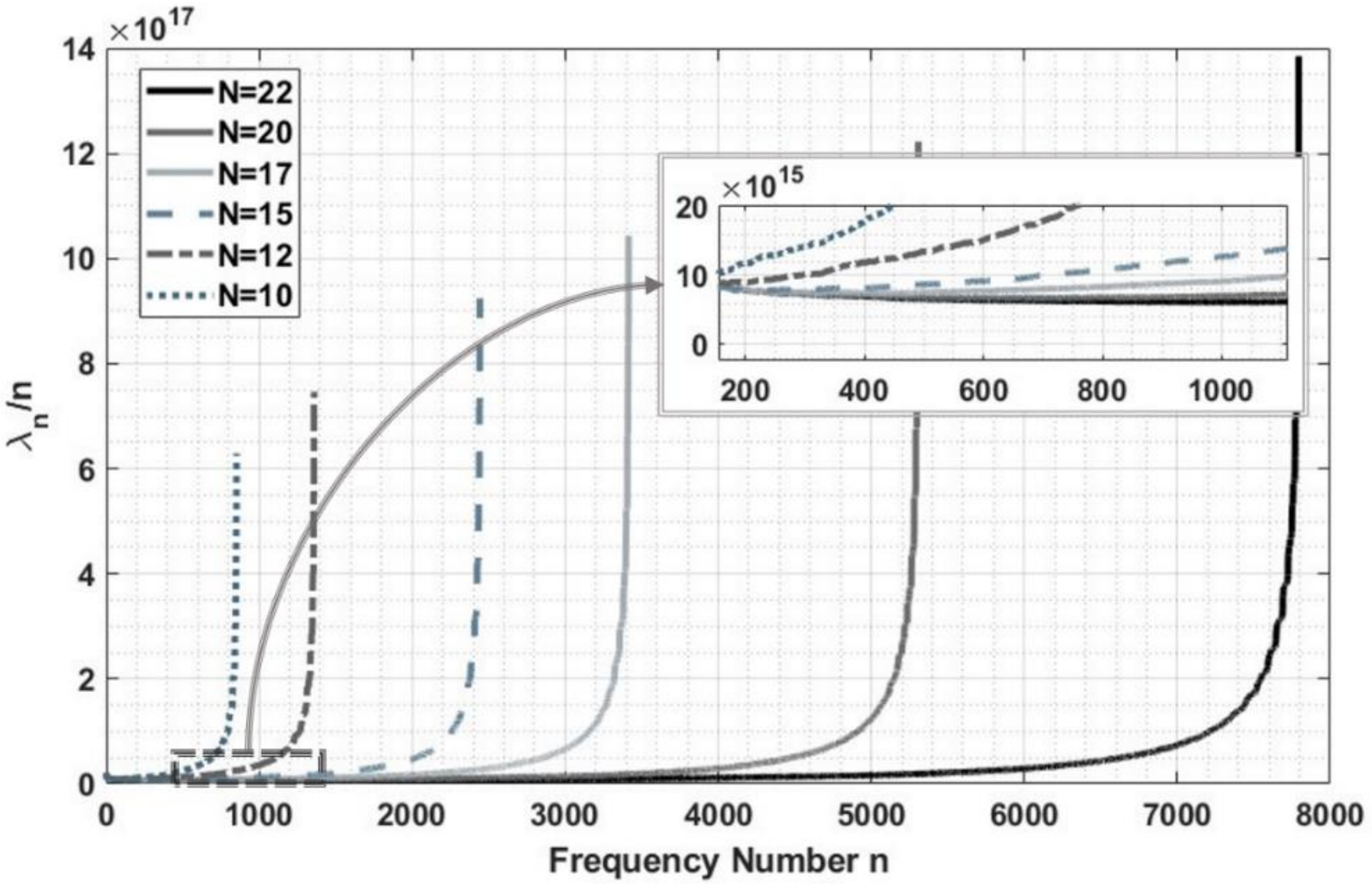

Figure 2: The plot of $\lambda n/n$ versus frequency number $n$ for different values of N. The data corresponds to the copper sample provided in Table 1, with dimensions $D_1$, $D_2$, and $D_3$ set to 1.5 mm, 2.5 mm, and 3.5 mm, respectively. As we increase N, the slope of the line approaches zero more rapidly and is sustained over a greater domain.

In section II-B-5, we have demonstrated the mathematical convergence for large values of *n*. While it was not feasible to run the numerical code for very large values of *n*, we were able to observe this convergence for *n* around one thousand, corresponding to N=22. The outcomes for the four tabulated materials, with the dimensions of the previous copper sample, are illustrated in Fig. 3.

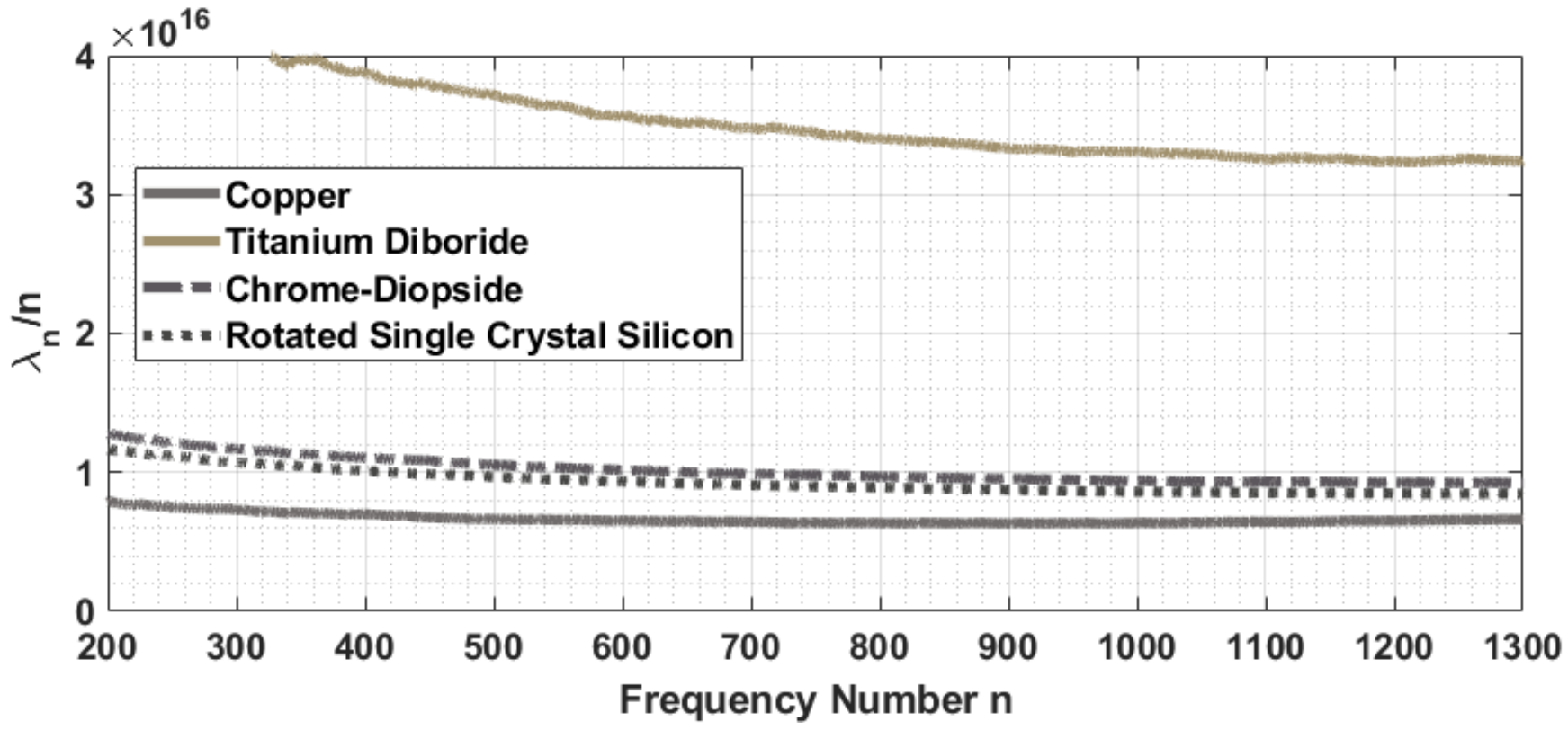

Figure 3: The plot of $\lambda n/n$ versus frequency number $n$ for different materials and N=22

## IV. CONCLUSIONS

In our investigation, we have examined the asymptotic behavior of eigenfrequencies, particularly how rapidly they approach infinity. Throughout this analysis, we have observed several factors related to the numerical limitations of the calculations and the impact of the diagonal elements on the limiting value. The existence of a limit for the asymptotes suggests that there is limited information available at very high frequencies. The asymptotic behavior was studied for four symmetry cases.

**REFERENCES**

1. Visscher, W.M., et al., *ON THE NORMAL-MODES OF FREE-VIBRATION OF INHOMOGENEOUS AND ANISOTROPIC ELASTIC OBJECTS.* Journal of the Acoustical Society of America, 1991. **90**(4): p. 2154-2162.
2. Migliori, A. and T.W. Darling, *Resonant ultrasound spectroscopy for materials studies and non-destructive testing.* Ultrasonics, 1996. **34**(2-5): p. 473-476.
3. Migliori, A., et al., *RESONANT ULTRASOUND SPECTROSCOPIC TECHNIQUES FOR MEASUREMENT OF THE ELASTIC-MODULI OF SOLIDS.* Physica B-Condensed Matter, 1993. **183**(1-2): p. 1-24.
4. Leisure, R.G. and F.A. Willis, *Resonant ultrasound spectroscopy.* Journal of Physics-Condensed Matter, 1997. **9**(28): p. 6001-6029.
5. Maynard, J.D., *THE USE OF PIEZOELECTRIC FILM AND ULTRASOUND RESONANCE TO DETERMINE THE COMPLETE ELASTIC TENSOR IN ONE MEASUREMENT.* Journal of the Acoustical Society of America, 1992. **91**(3): p. 1754-1762.
6. Demarest, H.H., *CUBE-RESONANCE METHOD TO DETERMINE ELASTIC CONSTANTS OF SOLIDS.* Journal of the Acoustical Society of America, 1971. **49**(3): p. 768-&.
7. Maynard, J., *Resonant ultrasound spectroscopy.* Physics Today, 1996. **49**(1): p. 26-31.
8. Sarrao, J.L., et al., *DETERMINATION OF THE CRYSTALLOGRAPHIC ORIENTATION OF A SINGLE-CRYSTAL USING RESONANT ULTRASOUND SPECTROSCOPY.* Review of Scientific Instruments, 1994. **65**(6): p. 2139-2140.
9. Farzbod, F. and D.H. Hurley, *Using Eigenmodes to Perform the Inverse Problem Associated with Resonant Ultrasound Spectroscopy.* Ieee Transactions on Ultrasonics Ferroelectrics and Frequency Control, 2012.
10. Barra, F., et al., *MEASURING DISLOCATION DENSITY IN ALUMINUM WITH RESONANT ULTRASOUND SPECTROSCOPY.* International Journal of Bifurcation and Chaos, 2009. **19**(10): p. 3561-3565.
11. Rossin, J., et al., *Assessment of grain structure evolution with resonant ultrasound spectroscopy in additively manufactured nickel alloys.* Materials Characterization, 2020. **167**: p. 110501.
12. McGuigan, S., et al., *Resonant ultrasound spectroscopy for quality control of geometrically complex additively manufactured components.* Additive Manufacturing, 2021. **39**: p. 101808.
13. Migliori, A. and J.D. Maynard, *Implementation of a modern resonant ultrasound spectroscopy system for the measurement of the elastic moduli of small solid specimens.* Review of Scientific Instruments, 2005. **76**(12).
14. Ogi, H., et al., *Contactless mode-selective resonance ultrasound spectroscopy: Electromagnetic acoustic resonance.* Journal of the Acoustical Society of America, 1999. **106**(2): p. 660-665.
15. Schley, R.S., et al., *Real-time measurement of material elastic properties in a high gamma irradiation environment.* Nuclear Technology, 2007. **159**(2): p. 202-207.
16. Reese, S.J., et al., *On the establishment of a method for characterization of material microstructure through laser-based resonant ultrasound spectroscopy.* Ieee Transactions on Ultrasonics Ferroelectrics and Frequency Control, 2008. **55**(4): p. 770-777.
17. Hurley, D.H., et al., *In situ laser-based resonant ultrasound measurements of microstructure mediated mechanical property evolution.* Journal of Applied Physics, 2010. **107**(6).
18. Amziane, A., et al., *Laser ultrasonics detection of an embedded crack in a composite spherical particle.* Ultrasonics, 2012. **52**(1): p. 39-46.
19. Hurley, D.H., S.J. Reese, and F. Farzbod, *Application of Laser-based Resonant Ultrasound Spectroscopy to Study Texture in Copper.* Journal of Applied Physics, 2012, to appear.
20. Nakamura, N., et al., *Determination of anisotropic elastic constants of superlattice thin films by resonant-ultrasound spectroscopy.* Journal of Applied Physics, 2005. **97**(1).
21. Ogi, H., et al., *Complete mode identification for resonance ultrasound spectroscopy.* Journal of the Acoustical Society of America, 2002. **112**(6): p. 2553-2557.

22. Sedlák, P., et al., *Determination of all 21 independent elastic coefficients of generally anisotropic solids by resonant ultrasound spectroscopy: benchmark examples.* Experimental Mechanics, 2014. **54**(6): p. 1073-1085.
23. Seiner, H., et al., *Application of ultrasonic methods to determine elastic anisotropy of polycrystalline copper processed by equal-channel angular pressing.* Acta Materialia, 2010. **58**(1): p. 235-247.
24. Wu, W., et al., *Measurement of Mechanical Properties of Hydrated Cement Paste Using Resonant Ultrasound Spectroscopy.* 2010.
25. Ostrovsky, L., et al., *Application of three-dimensional resonant acoustic spectroscopy method to rock and building materials.* Journal of the Acoustical Society of America, 2001. **110**(4): p. 1770-1777.
26. Kube, C.M., J. Gillespie, and M. Cherry, *Influence of residual stress and texture on the resonances of polycrystalline metalsa).* The Journal of the Acoustical Society of America, 2021. **150**(4): p. 2624-2634.
27. Kube, C.M., et al., *Resonant ultrasound spectroscopy for crystalline samples containing initial strain.* Journal of Applied Physics, 2022. **131**(22).
28. Kinney, J.H., et al., *Resonant ultrasound spectroscopy measurements of the elastic constants of human dentin.* Journal of Biomechanics, 2004. **37**(4): p. 437-441.
29. Bernard, S., Q. Grimal, and P. Laugier, *Accurate measurement of cortical bone elasticity tensor with resonant ultrasound spectroscopy.* Journal of the Mechanical Behavior of Biomedical Materials, 2013. **18**: p. 12-19.
30. Farzbod, F., *Resonant ultrasound spectroscopy for a sample with cantilever boundary condition using Rayleigh-Ritz method.* Journal of Applied Physics, 2013. **114**(2): p. 024902.
31. Visscher, W.M., et al., *On the normal modes of free vibration of inhomogeneous and anisotropic elastic objects.* Journal of the Acoustical Society of America, 1991. **90**(4;4 I;): p. 2154-2162.
32. Weber, H.J. and G.B. Arfken, *Essential mathematical methods for physicists, ISE.* 2003: Elsevier.
33. Meirovitch, L., *Fundamentals of vibrations.* 2001: McGrawHill, Singapore.
34. Meirovitch, L., *Elements of vibration analysis.* New York, McGraw-Hill Book Co., 1986, 574, 1986.
35. Trefethen, L.N. and D. Bau, *Numerical linear algebra.* Vol. 181. 1997: Society for Industrial and Applied Mathematics (SIAM).
36. Arfken, G., *Mathematical Methods for Physicists, 3rd edition.* 1985: Academic Press.
37. Sevigney, C.L., O.E. Scott-Emuakpor, and F. Farzbod, *Resonant Ultrasound Spectroscopy: Sensitivity Analysis for Anisotropic Materials With Hexagonal Symmetry.* Journal of Vibration and Acoustics, 2022. **144**(3).
38. Farzbod, F. and O.E. Scott-Emuakpor, *Resonant Ultrasound Spectroscopy: Sensitivity Analysis for Isotropic Materials and Anisotropic Materials With Cubic Symmetry.* Journal of Vibration and Acoustics, 2018. **141**(2): p. 021010-021010-10.
39. Farzbod, F. and D.H. Hurley, *Using eigenmodes to perform the inverse problem associated with resonant ultrasound spectroscopy.* IEEE transactions on ultrasonics, ferroelectrics, and frequency control, 2012. **59**(11).
40. Ledbetter, H. and A. Migliori, *A general elastic-anisotropy measure.* Journal of applied physics, 2006. **100**(6): p. 063516.
41. Isaak, D.G. and I. Ohno, *Elastic constants of chrome-diopside: application of resonant ultrasound spectroscopy to monoclinic single-crystals.* Physics and Chemistry of Minerals, 2003. **30**(7): p. 430-439.